\documentstyle[12pt]{article}
\begin{document}

\begin{center} {\LARGE \bf One-Dimensional Discrete Stark
\vskip0.3cm
 Hamiltonian and Resonance Scattering
\vskip0.3cm by Impurities
 }

\end{center}
\vskip 1cm

\noindent
{\bf Short title:} One-Dimensional Discrete Stark Hamiltonian...
\vskip1.5cm

\begin{center}
{\large  L.A.Dmitrieva$^{1,2}$, Yu.A.Kuperin$^{3}$ and Yu.B.Melnikov$^{1,3}$ }
\vskip 1cm
$^1$International Solvay Institutes for Physics and Chemistry, Campus Plain ULB
C.P.231, Boulevard du Triomphe, Brussels 1050, Belgium\\ {\it Tel.:
+32-2-6505536,
Fax: +32-2-6505028,}\\ {\it E-mail: melnikov@solvayins.ulb.ac.be}
\\
\vskip0.3cm

$^2$Department of Mathematical and Computational Physics, St.Petersburg State
University,   St.Petersburg 198904, Russia
\\
\vskip0.3cm

$^3$Laboratory of Complex Systems Theory, Institute for Physics, St.Petersburg
State
University,  St.Petersburg 198904, Russia\\ {\it Tel: +7-812-4284534, Fax:
+7-812-4284342,}\\ {\it E-mail: kuperin@cuper.niif.spb.su}

\end{center}

\newpage

\baselineskip=16pt

\centerline{\large Abstract}
\vskip0.3cm

A one-dimensional discrete Stark Hamiltonian with a continuous electric field
is
constructed by extension theory methods. In absence of  the
impurities the model is proved to be exactly solvable, the spectrum
is shown to be simple, continuous, filling the real axis; the
eigenfunctions, the resolvent and the spectral measure are
constructed explicitly.  For this (unperturbed) system the resonance
spectrum is shown to be empty.

The model considering impurity in a single node is also constructed using the
operator extension theory methods. The spectral analysis is performed and the
dispersion equation for the resolvent singularities is obtained. The resonance
spectrum is shown to contain infinite discrete set of resonances. One-to-one
correspondence  of the constructed Hamiltonian to some
Lee-Friedrichs model is
established.

\newpage

\baselineskip=18pt

\section{Introduction}

One-dimensional Stark type Hamiltonian on a line has been studied by many
authors [1-12]. Most attention has been paid to the resonance structure of this
system. The key ingredient of the models studied in [1-12] is absolutely
continuous
spectrum filling the whole real axis. It provides the possibility to apply
the powerful methods of the scattering theory to study the spectrum of
resonances.

However, it is well known [13,14] that discrete one-dimensional Stark
Hamiltonian
on a lattice has a discrete spectrum only. It means that
there is no "scattering states" even for unperturbed system. It
prevents to study the spectrum of resonances caused by the
perturbation of an ideal crystal lattice.

In the present paper we study a model for the motion of an electron on a
one-dimensional lattice in a homogeneous electric field and electron
resonance scattering by impurities treated as a perturbation.
The main idea of our
approach is to put the dynamical variables and equation of motion on a
spatial lattice, whereas the absolutely continuous spectrum is kept intact.

In order to construct a discrete Stark Hamiltonian with an
absolutely continous spectrum we treat the kinetic energy of an
electron in a lattice as the operator of the second difference
whereas the electric field is considered as continuous one filling
the intersite intervals. 
As it is shown in the paper the discrete Stark operator with the
continuous electric field proposed here has an absolutely
continuous spectrum in contrast to the discrete Stark operator
with the electric field located at the sites of the lattice
[13,14]. In our opinion the discrete Stark operator with
a continuous electric field seems to be more natural from the physical
point of view at least because the concept of a field needs the
continuaty by itself. On the other hand the suggested model
seems also to be more sound from the spectral point of view
because of the presence of a continuous spectrum and as a
consequence the presence of the propogating electronic waves in
the system.

Treating such Hamiltonian as an unpertubed operator we construct
the perturbed Hamiltonian which describes the interection of the
Stark electron with impurities.

In the present paper we consider the single impurity localized
at the site with the number $n=0$. We "switch on" the
interaction between the Stark electron and the impurity by the
extension theory methods [17,18]. In contrast to the ordinary
delta-like interaction our approach allows to take into account
the impurity internal degrees of freedom. The advantage is that 
the perturbed Stark operator describing the resonance scattering
by impurity leads to the exactly solvable model having at the
same time a reach set of resonances. We calculate the location
of resonances by perturbation theory methods in a weak coupling limit.  

We reduce also the proposed model to some
Lee-Friedrichs model [15,16] and treat the latter as a particular case
of models based on the extension theory [17,18]. This reduction
can be useful to study transition to chaos and
the problem of intrinsic irreversibility [19] for discrete Stark
Hamiltonians.

The paper is organized as follows. In Sec.2 we construct the
discrete Stark Hamiltonian with contionuous electric field and
perform its spectral analysis explicitly. In Sec.3 by means of
the extension theory methods we construct the perturbed
Hamiltonian describing the interaction of the Stark electron
with an impurity. In the same section we make the analytic
continuation of the resolvent bilinear form and in terms of this
continuation calculate the spectrum of resonances. In Seac.4 we
reduce our model to the Friedrichs-Lee one and discuss on this
basis the applicability of the generalized spectral
decompositions in connection with the intrinsic irreversibility
problem and chaotic regimes. 
\vskip1cm

\section{Unperturbed Hamiltonian}

In this section we construct a Hamiltonian describing
a chain of sites embedded in a continuous electric field
and study its spectral properties.

Let us consider the discrete Stark operator $H_d$ acting
in the Hilbert space ${\cal H}_d = l^2$ as follows
{\large
$$
(H_d\Psi)_n\,=\,-\,\frac{1}{(2\pi
a)^2}(\Psi_{n-1} + \Psi_{n+1}) + 2\pi \varepsilon a n \Psi_n
\eqno(1)
$$}
and the multiplication operator
{\large
$$
H_c\,=\, \varepsilon a y,\,\,\,\,\,\,\,\,\,\,y\,\in \,(-\pi,\pi)
\eqno(2)
$$ }
acting in the Hilbert space ${\cal H}_c\,=\,L^2(-\pi a,\pi a)$. 
Here $2 \pi a$ is the intersite distance and
$\varepsilon\,>\,0$ is the electric field parameter.

We introduce the Hamiltonian describing a chain of sites in an
electric field in the form
{\large
$$
H\,=\,H_d\times I_c\,+\,I_d\times H_c\,\,.
\eqno(3)
$$ }
Here $H_d$ and $H_c$ are given by Eqs.(1) and (2) respectively,
$I_d$ and $I_c$ are
the identity operators in ${\cal H}_d$ and ${\cal H}_c$. The notation $\times$
stands
for the operator tensor product. The operator $H$ acts in the space
{\large
$$
 {\cal H} = l^2({\bf Z};\,L^2[-\pi,\pi))\,\,.
\eqno(4)
$$ }
One can notice, that the potential
{\large
$$
U = 2\pi a\varepsilon n \times I_c + I_d\times \varepsilon a y =
\varepsilon a (2 \pi n + y)
$$ }
is a continuous electric field potential on a line. Indeed,
any point $\nu \in {\bf R}$ can be parametrized by the site
numder $n \in {\bf Z}$ and the point of the interval $y \in [-\pi
, \pi)$: $ \nu = a (2 \pi n + y)$. So in contrast to the
discrete Stark Hamoltonian $H_d$ our model describes a chain of
sites in a continuous electric field.

In what follows we call the operator $H$ as the unperturbed
operator. The pertubation will be introduced in the next section
as the impurity located in one of the sites. 

In order to describe spectral properties of the unperturbed operator
$H$ we need the following notations. By angle brackets
$\langle *,* \rangle_{\cal A}$ we denote the inner product in a
Hilbert space ${\cal A}$ and square brackets $[\alpha]$ stand for the
integer part of a real number $\alpha$. The integer-valued function
$M(\lambda)$ is defined as
{\large
$$
M(\lambda) \stackrel{\rm def}{=} \left[ \frac{\lambda}{2\pi a
\varepsilon} + \frac12 \right]\,\,;\,\,\,\,\lambda \in {\bf R}\,\,.
$$ }
We use the notation $J_{\nu}(z)$ for the Bessel functions of the
first type. The components of the vector $\widehat J^{(m)}
\in l^2$ are defined as follows:
$(\widehat J^{(m)})_n = J_{n-m}(\Theta)$, where
$\Theta = - (4\pi^3a^3\varepsilon)^{-1}$. The Heaviside step
function is denoted by $\theta(x)$.

{\bf Lemma 1.} {\it The spectrum $\sigma(H)$ of
the operator $H$ is simple,
absolutely continuous and fills the real axis {\bf R}.
The wave functions are distributions and are given by}
$$
\Psi_n(y,\lambda)\,=\,J_{n-M(\lambda)}(\Theta)\delta(\varepsilon a y -
\lambda
+ 2\pi M(\lambda) \varepsilon a)\,\,.
\eqno(5)
$$

{\it The spectral family (resolution of the identity) of the operator
$H$ are projections in the space ${\cal H}$ of the form} {\large $$
E_{\lambda}(y) = \sum_{m=-\infty}^{\infty}
< * , \widehat J^{(m)}>_{l^2} \widehat J^{(m)}
\theta (\lambda - 2 \pi m \varepsilon a - \varepsilon a y ) \,\,.
\eqno(6)
$$ }

\underline{Proof.} The structure of the operator $H$ leads to
the separation of variables and
therefore the spectral analysis of $H$ is
reduced to the spectral analysis of the operators $H_d$ and $H_c$.

Let us use the Fourier
trunsform in the space $l^2$, $F: l^2 \rightarrow L^2(-\pi, \pi)$:
{\large
$$
(Ff)(q)\,=\,\sum\limits_{n=-\infty}^{\infty} e^{inq}f_n
$$ }

In the Fourier representation the operator $H_d$ turns
into the operator
{\large
$$
FH_dF^{-1} = - 2 \pi a \varepsilon i \frac{d}{dq} -
\frac{2}{(2 \pi a)^2} {\rm cos} q ,
$$ }
acting in $L^2(-\pi ,\pi )$ whose
domain in the Sobolev space $H^1_2[-\pi ,\pi )$ is
determined by the periodicity condition
$(Ff)(-\pi ) = (Ff)(\pi )$.
Solving the eigenvalue problem
{\large
$$(FH_dF^{-1} - \lambda) (Ff)(q) = 0
$$ }
one can see that the spectrum of the operator $FH_dF^{-1}$ is
discrete and the eigenvalues are
{\large
$$\lambda_m = 2 \pi a \varepsilon m ,\,\,\,\,\, m = 0, \pm 1, \pm 2, ... ,
$$ }
whereas the correspondent eigenfunctions are
{\large
$$
(Ff)^{(m)}(q) = \exp \{i \Theta \sin q + i m q\} \,\,;\,\,\,\,
\Theta = - (4 \pi^3 a^3 \varepsilon)^{-1} \,\,.
$$ }

Using the inverse Fourier transform one obtains
{\large
$$
f^{(m)}_n = (2\pi )^{-1} \int_{-\pi}^{\pi} (Ff)^{(m)}(q)
e^{-inq} dq =
J_{n-m}(\Theta) \,\,.
$$ }
Here we have used the integral representation [29] of the Bessel
function $J_{\nu}(\Theta)$.

Now consider the operator $H_c$ in the Hilbert space
$L^2(-\pi, \pi)$. Its spectrum is absolutely continuous and fills the
interval $\zeta \in [-\pi \varepsilon a, \pi \varepsilon a]$.
The correspondent
continuous spectrum wave functions are distributions and have the form
of delta-functions $\psi_{\zeta}(y) = \delta(y - (\varepsilon a)^{-1}
\zeta)$.

Due to the separation of variables
the spectrum of the operator $H$ is the algebraic sum of spectra
of the operators $H_d$ and $ H_c$:

{\large
$$
\sigma(h) = \{z = z_1 + z_2: z_1 \in \sigma(H_d), z_2 \in
\sigma(H_c)\}\,.
$$ }
Let us note that the length of the single
spectral band of the operator $H_c$ exactly coinsides with the
distance between the neighbouring eigenvalues of the operator $H_d$.
Thus the spectrum of the operator $H$ fills the real axis.

Any point $\lambda \in {\bf R}$ from the
spectrum of $H$ can be represented in the
form
{\large
$$
\lambda = \lambda_m + \nu , \,\,\,\, m \in {\bf Z} ,
\,\,\, \nu \in [-\pi a \varepsilon  , \pi a\varepsilon ) .
$$ }
This representation corresponds to the energy distribution
between the "lattice" and "field" subsystems determined by the
operators $H_d$ and $H_c$ respectively. As
$\lambda_m = 2 \pi a \varepsilon m$ , $m = 0, \pm 1, \pm 2,
...$, the "mode number" $m$ for a given energy $\lambda$ is calculated as
an integer-valued function
$m = M(\lambda) = \left[ \frac{\lambda}{2\pi a\varepsilon} +
\frac12 \right]$.

Due to the separation of the variables the wave functions $\Psi_n(y)$
of the operator $H$ are product of the correspondent
eigenfunctions of the operators $H_d$ and $H_c$ and hence have
the form (5). 

Now we construct the spectral family $E_{\lambda}(y)$ of the operator $H$
and show that the quadratic form
{\large
$$
\eta(\lambda) = \langle E_{\lambda} \Phi , \Phi \rangle_{\cal H}
\eqno(7)
$$ }
is an absolutely continuous function of $\lambda$ for any $\Phi \in
{\cal H}$.
Firstly, let us show that the resolvent
$R(z) = (H - z)^{-1}$ is an
operator-valued matrix $R(z) = \{ R_{nn'}(z)\}$
with the entries
{\large
$$
R_{nn'}(y,z)\,=\,\sum\limits_{m=-\infty}^{\infty}
\frac{J_{n-m}(\Theta)J_{n'-m}(\Theta)}{\lambda_m+\varepsilon
ay-z}
\eqno(8)
$$ }
which acts as a multiplication operator with respect to the variable
$y$.  Indeed, the resolvent $R_d(z) = (H_d - z)^{-1}$ of the operator
$H_d$ in the space $l^2$ obviously has the matrix elements
{\large $$
\left(R_d(z)\right)_{nn'} = \sum_{m=-\infty}^{\infty}
\frac{J_{n-m}(\Theta) J_{n'-m}(\Theta)}{\lambda_m - z}\,\,,
\eqno(9)
$$ }
where $\lambda_m = 2\pi m\varepsilon a$ are the eigenvalues of the operator
$H_d$.
The resolvent $R_c(z) = (H_c - z)^{-1}$ of the operator $H_c$ in the space
$L^2(-\pi,\pi)$
is a multiplication operator
{\large
$$
R_c(z) \,*\, =\, \frac1{\varepsilon ay - z} \,*\,\,.
\eqno(10)
$$ }
Due to separation of variables
the resolvent $R(z) = (H - zI)^{-1}$ can be
calculated as a contour integral
{\large
$$
R(z) = (2\pi i)^{-1} \oint_{\gamma} R_c(\zeta)
R_d(z - \zeta) d\zeta =$$
$$= (2\pi i)^{-1} \int_{-\varepsilon a \pi}^{\varepsilon a \pi}
R_d(z-\zeta) [R_c(\zeta + i0) -
R_c(\zeta - i0)] d\zeta ,
$$ }
where the contour $\gamma$ encircles the spectrum of the operator
$H_c$.  On use of Eqs.(9),(10) the straightforward calculations
give
{\large
$$ R_{nn'}(y,z) = (2\pi i)^{-1} \int_{-\varepsilon
a\pi}^{\varepsilon a \pi} \sum\limits_{m=-\infty}^{\infty}
\frac{J_{n-m}(\Theta) J_{n'-m}(\Theta)} {\lambda_m-z+\zeta} \times $$
$$\times
\left( \frac1{\varepsilon ay - \zeta -i0} -
\frac1{\varepsilon ay - \zeta +i0} \right) d\zeta =$$
$$
= \sum_{m=-\infty}^{\infty} J_{n-m}(\Theta) J_{n'-m}(\Theta)
\int_{-\varepsilon a\pi}^{\varepsilon a \pi}
\frac{\delta(\varepsilon ay - \zeta)}{\lambda_m-z+\zeta}
d\zeta =$$
$$= \sum_{m=-\infty}^{\infty}
\frac{J_{n-m}(\Theta) J_{n'-m}(\Theta)}{\lambda_m - z + \varepsilon ay}
\,\,,
$$ }
what coincides with Eq.(8).

Resolution of the identity $E_{\lambda}$ of any selfadjoint operator
is related to the resolvent as follows [20]. If $(\alpha,\beta)
\subset {\bf R}$ is an open interval, then in the strong operator
topology {\large $$ E_{(\alpha,\beta)} = \lim_{\delta \downarrow 0}\,
\lim_{\epsilon \downarrow 0} \int_{\alpha+\delta}^{\beta-\delta}
\left(R(\lambda+i\epsilon) - R(\lambda - i\epsilon)\right)
d\lambda\,\,.
$$ }
Applying this formula to the resolvent $R(z)$ given by Eq.(8) one obtains
resolution of the identity of the operator $H$ in the form (6).

By means of Eq.(7) the function $\eta(\lambda)$ given by Eq.(8) takes the form
{\large
$$
\eta(\lambda) = \sum_{p=-\infty}^{\infty} \int_{-\pi}^{\pi} \langle \Phi,
\widehat J^{(p)}
\rangle_{ l^2}\, \langle \widehat J^{(p)}, \Phi \rangle_{l^2}\,
\theta(\lambda - \lambda_p - \varepsilon ay) dy\,\,.
$$ }
It is clear that if $\lambda \in [\lambda_m-\pi \varepsilon a, \lambda_m+\pi
\varepsilon a]$,
$m \in {\bf Z}$, then
{\large
$$
\eta(\lambda) = \sum_{p=-\infty}^{m-1} \int_{-\pi}^{\pi}
\langle \Phi, \widehat J^{(p)}
\rangle_{l^2}\, \langle \widehat J^{(p)}, \Phi \rangle_{l^2}\,dy \,+
$$
$$
+\,
\int_{-\pi}^{(\lambda-\lambda_m)/\varepsilon a}
\langle \Phi, \widehat J^{(m)}
\rangle_{l^2}\, \langle \widehat J^{(m)}, \Phi \rangle_{l^2}\,dy\,\,.
$$ }
Consequently, $\eta(\lambda_m+\pi \varepsilon a-0) =
\eta(\lambda_{m+1}-\pi \varepsilon a+0)$. Hence the function $\eta(\lambda)$
and, consequently,
the spectrum of the operator $H$, is absolutely continuous.

Lemma is proved.

\section{Perturbed Hamiltonian}

In this section we assume that the considered above chain of nodes
embedded in the electric field has an impurity.
Namely we suppose that the electron dynamics governed by the
Hamiltonian $H$ is perturbed by the additional interaction
between the electron and an impurity located at the single site
with the number $n=0$. We are going to construct this
interaction by means of the extension theory methods [17,18].
Namely we suppose that the impurity has an internal structure.
The dynamics of this internal structure is given by a
self-adjoint operator $H_i$ acting in an auxilary Hilbert space
${\cal H}_i$. Then following the extension theory ideology
we consider the extended space
{\large
$$
\widehat {\cal H} = l^2({\bf Z}, L^2(-\pi,\pi)\oplus{\cal H}_i)
= {\cal H} \oplus l^2({\bf Z}, {\cal H}_i)\,\,
$$ }
as the state space for the system with interaction.

Here we make the simplest choice  ${\cal H}_i = {\bf
C}$. Then
{\large
$$
\widehat {\cal H} = {\cal H} \oplus l^2
$$ }
and the self-adjoint operator $H_i$ acting in the auxilary space
${\cal H}_i$
is just the operator operator of multiplication by a real number $\mu
\in {\bf R}$,
{\large
$$ H_i \stackrel{{\rm def}}{=} \mu *\,\,.
$$ }

Let us embed the operator $H$,  in
the space $\widehat {\cal H}$ as follows
{\large
$$
H \rightarrow \widehat H =
H_d \times (I_c \oplus I_i) + I_d \times (H_c \oplus H_i) = $$ $$ =
(H_d \times I_c + I_d \times H_c) \oplus (H_d \times I_i + I_d \times
H_i) = $$ $$ =\left( \begin{tabular}{c c}
$H$ & $0$\\
$0$ & $H_d \times I_i + I_d \times H_i$\\
\end{tabular} \right) \,\,,
\eqno(11)
$$ }
where $I_i$ stands for the identity operator in ${\bf C}$. 
The diagonal structure of this operator means that the embedding
does not lead to any interaction between the Stark electron and
the impurity.

In order to "switch on" the interaction one can add to the diagonal
operator-valued matrix (11) an off-diagonal self-adjoint
operator $V$ :
{\large $$
\widehat H_B = \widehat H_0 + V\,\,,
\,\,\,\,V = \left(
\begin{tabular}{c c}
$0$ & $B$\\
$B^+$ & $0$\\
\end{tabular}
\right)\,\,,
\eqno(12)
$$ }
where $B$ is a bounded operator acting from
$l^2$ to ${\cal H}$ and $B^+$ is its
adjoint. Obviously $\widehat H_B$ is self-adjoint operator with
the domain of $\widehat H$.

As we wish to have an
additional interaction only with a single site (say, with $n = 0$),
the operator $B$ should vanish in the orthogonal complement to the
linear span ${\cal L}\{\chi\}$  of the vector
$\chi = (...,0,0,1,0,0,...)^T \in l_2$, ($\chi_n = \delta_{0n}$),
i.e.
{\large
$$
B|_{l^2 \ominus {\cal L}\{\chi\}} = 0\,\,.
$$ }
This gives us the form\footnote{One can consider impurities localized
in any finite number $N$ of
nodes. It
means that the operator $B$ should have non-trivial components
in $N$-dimensional subspace of the space $l^2$, which
makes the algebraic structure of the result more complicated but
do not lead to any essentialy new spectral effects.}
of the operator $B$:
{\large
$$
B:\,f \mapsto \beta \langle f, \chi \rangle_{l^2} \widehat \chi\,\,.
\eqno(13)
$$ }
Here $\beta \in {\bf R}$ is a coupling constant and
$\widehat \chi = \chi \cdot \varphi (y) \in {\cal H}$.
We consider here the interaction the interaction which does not
depend on the field variable
$y$, so it is reasonable to suppose that the function
$\varphi(y) \in L^2[-\pi,\pi)$ is a constant, $\varphi (y) \equiv 1$,
and
{\large
$$
\widehat \chi = \chi \cdot {\bf 1}(y) \in {\cal H}\,\,.
$$ }
The adjoint operator acts as
{\large
$$
B^+:\, \Psi \mapsto \beta \langle \Psi, \widehat \chi \rangle_{\cal
H} \chi\,\,.
\eqno(14)
$$ }

To study the spectral properties of the operator $\widehat H_B$
we consider the
spectral problem
{\large
$$
\widehat H_B \left(
\begin{tabular}{c}
$\Psi(y)$ \\
$f$\\
\end{tabular} \right) = z
\left(
\begin{tabular}{c}
$\Psi(y)$ \\
$f$\\
\end{tabular} \right) \,\,,\,\,\,\,\Psi \in {\cal H}\,\,,\,\,\,f \in l_2\,\,,
\eqno(15)
$$ }
and eliminate the "impurity channel" variable
{\large
$$
f = - \beta \langle \Psi, \widehat \chi\rangle_{\cal H}\, R_d(z-\mu) \chi\,\,,
$$ }
where $R_d(z) = (H_d - z)^{-1}$. This leads to the effective equation
{\large
$$
\left( H + W_{11}(z) -z \right) \Psi = 0
\eqno(16)
$$ }
in the space ${\cal H}$ with the energy-dependent interaction [17,18]
{\large
$$
W_{11}(z) * = - B R_d(z-\mu) B^+ * =
 - \beta^2 \langle *, \widehat
\chi \rangle_{\cal H} \langle R_d(z-\mu) \chi, \chi \rangle_{l_2}
\widehat \chi\,\,.
$$ }
Using Eq.(16) one can write [17,18] the Lippmann-Schwinger equation
for the block $\widehat R_{11}(z) = (H + W_{11}(z) - z)^{-1}$ of the
resolvent $(\widehat H_B - z)^{-1} = R_B(z) = \{\widehat R_{ij}\}_{i,j=1}^2$ :
{\large
$$
\widehat R_{11}(z) = R(z) - R(z) W_{11}(z) \widehat R_{11}(z) \,\,.
\eqno(17)
$$ }
This equation has an exact solution
{\large
$$
\widehat R_{11}(z) = R(z) +
\frac{\beta^2}{Q(z)} \langle R_0(z)*, \widehat \chi, \widehat \chi\rangle_{\cal
H}\,
\langle R_d(z-\mu)\chi, \chi \rangle_{l_2}
\eqno(18)
$$ }
where Krein's determinant $Q(z)$ [17,18] is given by
{\large
$$
Q(z) = 1 - \beta^2 \langle R(z) \widehat \chi, \widehat \chi \rangle_{\cal H}
\langle R_d(z-\mu) \chi, \chi \rangle_{l_2}\,\,.
\eqno(19)
$$ }

Similary, one can eliminate from Eq.(15) the variable
{\large
$$
\Psi(y) = - R(z) B f = - \beta \langle f, \chi \rangle_{l_2}
R(z) \widehat \chi (y).
$$ }
In this case one obtains an effective equation
{\large
$$
\left( H_d + \mu - z + W_{22}(z) \right) f = 0
$$ }
with the energy-dependent interection
{\large
$$
W_{22}(z) \,* = - B^+ R(z) B =
- \beta^2 \langle *, \chi \rangle_{l_2}\,
\langle R(z) \widehat \chi, \widehat \chi
\rangle_{\cal H}\, \chi\,\,.
$$ }
The correspondent Lippmann-Schwinger equation has again an exact solution
{\large
$$
\widehat R_{22}(z) = R_d(z-\mu) + \frac{\beta^2}{Q(z)}
\langle R(z) \widehat \chi, \widehat \chi \rangle_{\cal H}
\, \langle R_d(z-\mu) *, \chi \rangle_{l^2}\, R_d(z-\mu) \chi\,\,.
\eqno(20)
$$ }

The operators $\widehat R_{11}(z)$ and $\widehat R_{22}(z)$ are
the diagonal elements of
the resolvent of the operator $\widehat H_B$,
{\large
$$
(\widehat H_B - z)^{-1} = \widehat R_B(z) =
\left(
\begin{tabular}{c c}
$\widehat R_{11}(z)$ & $\widehat R_{12}(z)$\\
$\widehat R_{21}(z)$ & $\widehat R_{22}(z)$\\
\end{tabular} \right)\,\,.
$$ }
It remains to reconstruct the off-diagonal elements $\widehat R_{12}(z)$,
$\widehat R_{21}(z)$.
To this end let us consider the Lippmann-Schwinger equation
for the resolvent $\widehat R_B(z)$ of the operator $\widehat H_B$:
{\large
$$
\widehat R_B(z) = \widehat R(z) - \widehat R(z) V \widehat R_B(z)\,\,.
\eqno(21) 
$$}
Here
{\large
$$
\widehat R(z) = \left(
\begin{tabular}{c c}
$R(z)$ & $0$\\
$0$ & $ R_d(z-\mu)$ \\
\end{tabular} \right)
$$ }
is the resolvent of the operator $\widehat H$ and
$V$ is the perturbation.
On the basis of the Lippman-Schwinger equation (21)
the off-diagonal elements of the operator-valued matrix
$\widehat R_B(z)$ are eaisely expressed through the diagonal
ones as follows
{\large
$$
\widehat R_{12}(z) = - R(z) B \widehat R_{22}(z)\,\,,
\eqno(22)
$$
$$
\widehat R_{21}(z) = - R_d(z-\mu) B^+ \widehat R_{11}(z)\,\,.
\eqno(23)
$$ }

The analysis of the analytical structure of the constructed
resolvent $\widehat R_B(z)$ leads to the following conclusions.
The resolvent $\widehat R_B(z)$ is an analytic operator-valued
function in the upper (${\rm Im} z > 0$) and lower (${\rm Im} z < 0$)
halfplanes. It has a jump on the continuous spectrum of the
operator $\widehat H_B$ which coincides with the real axis ${\bf R}$.
Such analytic properties are the direct consequence of the same
analytic properties of the resolvent $R(z)$ and of the fact that the poles
of the resolvent $R_d(z-\mu)$, namely the points
$z_m = 2\pi \varepsilon am + \mu$, $m \in {\bf Z}$, are cancelled in each
matrix
element of $\widehat R_B(z)$. This cancellation can be
easely checked on use of the explicit formulae (18), (20), (22),
(23) for the matrix elements.

It should be noted that the resolvent $\widehat R(z)$ of the
unperturbed imbeded operator
$\widehat H$  besides the jump on the real axis has also the poles
at the points $z_m = \lambda_m + \mu$, $m \in {\bf Z}$. So the
eigenvalues of the operator $\widehat H$ are embedded into the
continuous spectrum. Thus we have shown that the adding of the
impurity destroyes these eigenvalues.
In what follows we show that under the perturbation $V$
eigenvalues convert into an infinite set of resonances.

We shall define resonances as the poles of the
analytic continuation of the
quadratic form (see [21] and references therein):
{\large
$$
\tau(z) = \langle \widehat R(z) u, v \rangle_{\widehat{\cal H}}\,\,.
\eqno(24)
$$ }
Here the vectors
{\large
$$
u = \left(
\begin{tabular}{c}
$u_1$\\
$u_2$\\
\end{tabular} \right)\,\,,\,\,\,\,
v = \left(
\begin{tabular}{c}
$v_1$\\
$v_2$\\
\end{tabular} \right)
$$ }
are appropriate elements form the space $\widehat{\cal H} = {\cal H}
\oplus l^2$ ($u_1, v_1 \in {\cal H}$ and $u_2, v_2 \in l^2$) to be
specified below.

Let us denote as $\widehat{\cal H}_A$ the subset of the space
$\widehat{\cal H}$ consisting of elements $u$ whose components $u_1(y)$
admit analytic continuation into the strip $|{\rm Re}\, y| < \pi$. The space
$\widehat{\cal H}_A$ is dense in $\widehat{\cal H}$. Indeed, since
$u_1(y) \in L_2(-\pi, \pi) \times l^2$, one can take polynomials as a dense
subset in $L_2(-\pi, \pi)$ which elements admit the above analytic
continuation.
Let us introduce the notation $\lambda_m^{\pm} = \lambda_m \pm \pi
\varepsilon a$. The following statement is valid

{\bf Lemma 2}. {\it Let $u, v \in \widehat{\cal H}_A$. Then the form $\tau(z)$
admits meromorphic continuation from above (below) to below (above) in any
strip $S_m = \{z: \lambda_m^- < {\rm Re}\, z < \lambda_m^+\}$ and its poles
coincide with the zeros of the correspondent continuation of the Krein's
determinant $Q(z)$.}

\underline{Proof.} Let us rewrite the quadratic form (24) as follows
{\large
$$
\tau(z) = \langle R_{11}(z) u_1 + R_{12}(z) u_2, v_1 \rangle_{\cal H} +
\langle R_{21}(z) u_1 + R_{22}(z) u_2, v_2 \rangle_{l^2}\,\,.
\eqno(25)
$$ }
Since the proof for all terms in Eq.(25) is similar, we show only that
$\langle R_{11}(z) u_1, v_1 \rangle_{\cal H}$ admits analytic continuation into
the strip $\lambda_m^- < {\rm Re}\, z < \lambda_m^+$. Using Eq.(18) one
obtains
{\large
$$
\langle R_{11}(z) u_1, v_1 \rangle_{\cal H} = \langle R(z) u_1, v_1
\rangle_{\cal H} +
$$
$$
+ \frac{\beta^2}{Q(z)} \langle R_d(z-\mu) \chi, \chi \rangle_{l^2}
\langle R(z) u_1, \widehat \chi \rangle_{\cal H}\,
\langle R(z) \widehat \chi, v_1 \rangle_{\cal H} \,\,.
\eqno(26)
$$ }
First let us show that the first term in the right hand side of
Eq.(26) admits analytic continuation into the strip $S_m$ for any
$m$. On use of Eq.(9) we have
{\large $$ \langle R(z) u_1, v_1
\rangle_{\cal H} = \sum_{m n n'} J_{n-m}(\Theta) J_{n'-m}(\Theta)
\varphi^{nn'}_m(z)\,\,, \eqno(27) $$ }
where
{\large $$
\varphi^{nn'}_m(z) \stackrel{{\rm def}}{=} \int_{-\pi}^{\pi}
\frac{\left(u_1(y)\right)_{n'} \left(v_1(y)\right)^*_n}{\lambda_m +
\varepsilon ay - z} dy\,\,.
\eqno(28)
$$ }
Since the function $\varphi^{nn'}_m(z)$ is given by the Cauchy type
integral, it is analytic on the complex plane of variable $z$
except the interval $[\lambda_m^-, \lambda_m^+]$. Let us introduce the
following functions on the strip $ \lambda_m^- < |{\rm Re}\, z| <
\lambda_m^+$ :
{\large
$$
^-\varphi_m^{nn'}(z) \stackrel{{\rm def}}{=} \left\{
\begin{tabular}{l r}
$\varphi_m^{nn'}(z)$\,\,,&\,\,\,\,${\rm Im}\, z > 0$\\
$\varphi_m^{nn'}(z) + \frac{2\pi i}{\varepsilon a}\,
h_{nn'}\left(\frac{z-\lambda_m}{\varepsilon a}\right)$\,\,,&
\,\,\,\,${\rm Im}\, z < 0$\\
\end{tabular}
\right.
\eqno(29)
$$ }
and
{\large
$$ ^+\varphi_m^{nn'}(z) \stackrel{{\rm def}}{=} \left\{
\begin{tabular}{l r}
$\varphi_m^{nn'}(z) - \frac{2\pi i}{\varepsilon a}\,
h_{nn'}\left(\frac{z-\lambda_m}{\varepsilon a}\right)$\,\,,&
\,\,\,\,${\rm Im}\, z > 0$\\
$\varphi_m^{nn'}(z)$\,\,,&\,\,\,\,${\rm Im}\, z < 0$\\
\end{tabular}
\right.\,\,,
\eqno(30)
$$ }
where
{\large
$$
h_{nn'}(z) \stackrel{{\rm def}}{=} \left(u_1(z)\right)_{n'}
 \left(v_1(z)\right)^*_n
$$ }
is the analytic continuation of the Cauchy type integral density (28)
into the strip $S_m$.

One can check that the functions $^{\pm}\varphi^{nn'}_m(z)$ are analytic
in the strip $S_m$. To this end it is enough to show that
{\large
$$
^{\pm}\varphi^{nn'}_m(\lambda + i0) = {^{\pm}\varphi^{nn'}_m}(\lambda - i0)
$$ }
for any $\lambda \in {\bf R}$. This relation follows directly from the
limit values of the Cauchy type integral $\varphi_m^{nn'}(z)$ on the
interval $(\lambda^-_m, \lambda_m^+)$,
{\large
$$
\varphi_m^{nn'}(\lambda \pm i0) =
\pm \frac{\pi i}{\varepsilon a}\,
h_{nn'}\left(\frac{\lambda-\lambda_m}{\varepsilon
a}\right) + {\rm P.V.}\int_{-\pi}^{\pi} \frac{h_{nn'}(y)}{\lambda_m
+\varepsilon ay - \lambda} dy\,\,.
\eqno(31)
$$ }
Thus the function $^-\varphi_m^{nn'}(z)$ is the analytic continuation
of the function $\varphi_m^{nn'}(z)$ from above to below and the function
$^+\varphi_m^{nn'}(z)$ is the analytic continuation of the function
$\varphi_m^{nn'}(z)$ from below to above in the strip $S_m$. Hence the form
$\langle R(z) u_1, v_1 \rangle_{\cal H}$
given by the series (27) admits analytic continuation from
above to below and vice versa in each strip $S_m$. From Eqs.(29) --
(31) it follows that these continuations through the interval
$(\lambda^-_m, \lambda^+_m)$ have the form {\large $$ \langle R(z)
u_1, v_1 \rangle^{\mp}_{\cal H} = \sum_{lnn'} J_{n-l}(\Theta)
J_{n'-l}(\Theta) \varphi_l^{nn'}(z) \pm $$ $$ \pm \frac{2\pi
i}{\varepsilon a} \sum_{nn'} J_{n-m}(\Theta) J_{n'-m}(\Theta)
h_{nn'}\left(\frac{z - \lambda_m}{\varepsilon a}\right) \,\,.
\eqno(32)
$$ }
These functions are analytic in the strip $S_m$. They can also be considered as
the
fixed branches of the functions given by the same formulae (27) and
(28) on the complex plane of variable $z$ with two cuts along the
rays $(-\infty, \lambda^-_m]$ and $[\lambda_m^+, \infty)$.

Let us consider now the second term in the right hand side of
Eq.(26). The analytic continuation of the factors $\langle R(z) u_1,
\widehat \chi \rangle_{\cal H}$ and $\langle R(z) \widehat \chi, v_1
\rangle_{\cal H}$ is a consequence of the analytic continuation
proven above for the form $\langle R(z) u_1, v_1 \rangle_{\cal H}$.
The factor $\langle R_d(z-\mu) \chi, \chi \rangle_{l^2}$ is a
meromorphic function with the poles at the points $\lambda_m+\mu$.
However, one can check that these poles are cancelled by the same
poles of the Krein's determinant $Q(z)$ (and its analytic
continuations $Q^{\pm}(z)$ as well).

The last factor to be considered is the Krein's determinant $Q(z)$.
Using Eqs.(9), (19) and (28) one can rewrite it as
{\large
$$
Q(z) = 1 - \frac{\beta^2}{\varepsilon a}\, \left(
\sum_{p=-\infty}^{\infty} \frac{J_p^2(\Theta)}{\lambda_p-z+\mu}\right)
\left( \sum_{p=-\infty}^{\infty} J_p^2(\Theta) {\rm ln}\, \left(
\frac{\lambda_p^+-z}{\lambda_p^--z}\right) \right)\,\,.
\eqno(33)
$$ }
Here the branch of the logarithm is choosen such that
{\large
$$
\left.
{\rm ln}\, \left(
\frac{\lambda_p^+-z}{\lambda_p^--z}\right) \right|_{z=\lambda+i0} =
i\pi +
{\rm ln}\, \left(
\frac{\lambda_p^+-\lambda}{\lambda-\lambda_p^-}\right)\,\,,\,\,\,\,
\lambda \in (\lambda_p^-, \lambda_p^+)\,\,.
\eqno(34)
$$ }
One can check that the analytic continuation of the function $Q(z)$
into the strip $S_m$
from above (below) to below (above) is given by the function
$Q^-(z)$ ($Q^+(z)$) :
{\large
$$
Q_m^-(z) \stackrel{{\rm def}}{=}
\left\{
\begin{tabular}{l r}
$Q(z)$\,\,,&\,\,\,\,${\rm Im}\, z > 0$\\
$Q(z) - \frac{2\pi i \beta^2}{\varepsilon a}\, J^2_m(\Theta)\,
\sum_{p=-\infty}^{\infty}
\frac{J_p^2(\Theta)}{\lambda_p-z+\mu}$\,\,,&
\,\,\,\,${\rm Im}\, z < 0$\\
\end{tabular} \right.\,\,,
\eqno(35)
 $$ }
{\large
$$ Q_m^+(z) \stackrel{{\rm def}}{=}
\left\{
\begin{tabular}{l r}
$Q(z) + \frac{2\pi i \beta^2}{\varepsilon a}\, J_m^2(\Theta)\,
\sum_{p=-\infty}^{\infty}
\frac{J_p^2(\Theta)}{\lambda_p-z+\mu}$\,\,,&
\,\,\,\,${\rm Im}\, z > 0$\\
$Q(z)$\,\,,&\,\,\,\,${\rm Im}\, z < 0$\\
\end{tabular} \right.\,\,.
\eqno(36)
 $$ }

Therefore the form $\langle R_{11}(z) u_1, v_1 \rangle_{\cal H}$ can be
continued
in each strip $S_m$ as a meromorphic function and the only poles can be given
by the zeroes of the Krein's determinant continuations $Q^{\pm}(z)$.

The proof of the statement of Lemma 2
for all other terms of the form $\tau(z)$ is similar. This acomplishes
the proof of the lemma.

From the above proof one can see that if the functions
$Q^-_m(z)$ and $Q_m^+(z)$ have zeroes $z_m^-$ and $z_m^+$ in the half-strips
$S_m^- = S_m \cap \{z: {\rm Im}\, z < 0\}$ and
$S_m^+ = S_m \cap \{z: {\rm Im}\, z > 0\}$ respectively, these
points are resonances.

To analyze the localization of the resonances let us consider the
weak coupling limit $\beta \ll 1$.

The following statement is valid

{\bf Theorem 1.} {\it The quadratic form
$\tau(z) = \langle \widehat R_B(z) u, v \rangle_{\widehat {\cal H}}$
for any $u, v \in \widehat {\cal H}$ is
an analytic function in upper (${\rm Im} z > 0$) and lower (${\rm Im}
z < 0$) halfplanes and has a jump on the real axis.  The meromorphic
continuations $\tau_m^{\pm}(z)$ of $\tau(z)$ into the strip $S_m$
have a set of poles
(resonances) in the upper and lower halfstrips respectively which
coincide with zeros of continuations $Q^{\pm}_m(z)$ of the Krein's determinant
$Q(z)$. In the week-coupling limit ($\beta\ll
1$) there are at least one pole of
$\tau^+_m(z)$ ($\tau^-_m(z)$) in the
lower (upper) halfstrip for every $m \in {\bf Z}$ given by }
{\large
$$
z^{\pm}_m\,\mathop{=}\limits_{\beta\ll
1}\,\lambda_{M(\lambda_m-\mu)} + \mu -
$$
$$
- \frac{\beta^2}{\varepsilon a} \,J^2_{M(\lambda_m-\mu)}(\Theta)
\sum_{p=-\infty}^{\infty} J^2_p(\Theta)\, \ln \left|
\frac{\lambda^+_p - \lambda_{M(\lambda_m-\mu)} - \mu}{\lambda^-_p -
\lambda_{M(\lambda_m-\mu)} - \mu} \right| -
$$
$$
- 3i\pi \frac{\beta^2}{\varepsilon a}\, J^2_{M(\lambda_m-\mu)}(\Theta)
J^2_m(\Theta) + O(\beta^4)\,\,.
\eqno(37)
$$ }
{\it Here the Bessel function subindex
$M(\lambda_m - \mu)$ is given by the integer-valued function
$M(\lambda) = \left[ \frac{\lambda}
{2\pi \varepsilon a} + \frac12 \right]$ and $\lambda_m^{\pm} = \pi \varepsilon
a(2m \pm 1)$. }

\underline{Proof.}
Analyticity of the function $\tau(z)$in ${\bf C}\backslash{\bf R}$ is a
straightforward consequence of the analytic properties of the resolvent
$\widehat R(z)$ of the selfadjoint operator $\widehat H_B$.

Lemma 2 implies that singularities of the continuations $\tau_m^{\pm}(z)$
 coincide with the
zeros of the
continuations $Q_m^{\pm}(z)$ of Krein's determinant $Q(z)$.
In the upper halfstrip $Q^+_m(z) = Q(z)$ and has no zeros, while
in the lower halfstrip $S^-_m$ due to Eqs.(33) and (35)
zeros of $Q_m^+(z)$ are given by the
roots of the equation
{\large
$$
\frac{\beta^2}{\varepsilon a}\, \sum_{n=-\infty}^{\infty}
\frac{J^2_n(\Theta)}{\lambda_n - z +\mu} \left(
\sum_{p=-\infty}^{\infty} J^2_p(\Theta) \ln
\left(\frac{\lambda_p^+-z}{\lambda^-_p-z}\right)
\, +\, 2i\pi J^2_m(\Theta) \right) \, = \, 1\,\,.
\eqno(38)
$$ }
 Let us show that in the weak coupling limit
$\beta \ll 1$ at least one pole of the resolvent
$(H_d-z+\mu)^{-1}$ generates a root of Eq.(38), and, consequently a
pole of $\tau^+_m(z)$ in the
lower halfplane (resonance). Considering
$\tau_m^-(z)$ one should just replace lower halfplane by the upper
one and vice versa.

We assume $\mu \ne n + 1/2$, $n \in {\bf Z}$, and choose the index
$m' \in {\bf Z}$ such that $\lambda_{m'} + \mu \in (\lambda_m^-,
\lambda_m^+)$. That means
{\large $$ m' = M(\lambda_m - \mu)\,\,.
\eqno(39)
$$ }
Multiplying both hand sides of Eq.(38) by the factor
$(\lambda_{m'} - z - \mu)$ and using the expansion in powers of
$\beta^2$ near $\beta = 0$ we find the root
{\large
$$
z^+_m= \lambda_{m'} + \mu -
$$
$$
- \frac{\beta^2}{\varepsilon a} \, J^2_{m'}(\Theta)\,
\left(\sum_{p=-\infty}^{\infty} J^2_p(\Theta)\,
\ln
\left(\frac{\lambda_p^+-\lambda_{m'}-\mu}{\lambda_p^--\lambda_{m'}-\mu}\right)
\, + \, 2i\pi J^2_m(\Theta)\right)\,\,.
\eqno(40)
$$ }
The argument of the logarithm in the right hand side of Eq.(40) is
negative iff $\lambda_p^- - \mu < \lambda_{m'} < \lambda_p^+ -
\mu$, i.e. when $p = M(\lambda_{s(m)}+\mu)$. One use of Eq.(39) we
get that it means $p = m$.  Finally, using Eqs.(34), (39) and (40) we
get the resonance in the lower halfplane given by Eq.(37).
Calculations of the resonances $z_m^-$ in the upper halfplane
are obviously similar. Theorem is proved.

\section{Lee-Friedrichs representation}

The Hamiltonian $\widehat H$ acts in the Hilbert space
$\widehat{\cal H}$ wich can be represented as the orthogonal sum
$\widehat{\cal H} = l^2({\bf Z}, L^2[-\pi,\pi)) \oplus l^2$.
In accordance with this representation let us introduce
the generalized Friedrichs states
$|\omega>$ and $|s>$ [16,19,22-24] as follows
{\large
$$
|\omega>\,=\,
\left({
\begin{tabular}{c}
$|\tilde\omega>$\\
0
\end{tabular}
}\right),
$$ }
where
$|\tilde\omega> =
\widehat J^{(M(\omega))} \delta(y+2\pi
M(\omega)-\frac{\omega}{\varepsilon a})$,
{\large
$$
|s>\,=\,
\left({
\begin{tabular}{c}
0 \\
$|\tilde s>$  \\
\end{tabular}
}\right),
$$ }
and
$ |\tilde s \rangle = \widehat J^{(s)} = \{J_{n-s}\}_{n\in {\bf Z}} \in l^2$.
Then one can chek that the perturbed
Hamiltonian $\widehat{H_B}$ can be written in the
Lee-Friedrichs representation
{\large
$$
\widehat{H_B}\,=\,2\pi a\varepsilon\sum_{s=-\infty}^{\infty} s\,
|s><s|\,+\,\int_{-\infty}^\infty
d\omega\,\omega\,
|\omega><\omega|\,+
$$
$$
+\,\sum_{s=-\infty}^\infty
\int_{-\infty}^\infty d\omega\, v_s(\omega)\,\left( |\omega><s|\,+\,
|s><\omega| \right)\, ,
\eqno(41)
$$ }
where the spectral density $v_s(\omega)$ reads as
{\large
$$
v_s(\omega)\,=\,\beta
J_{-M(\omega)}(\Theta)J_{-s}(\Theta),\,\,\,\,\omega\in {\bf R}.
$$ }

The representation (41) shows that our model can be effectively
studied using methods developed in various papers [16,22,23,25-27] for
the Friedrichs model. The main interest in this model is related
to consideration of resonance states assosiated with resonance
poles of the extended resolvents, their interpretation as
decaying states and construction of generalized spectral
decomposition [9]. However, one should notice that the constructed
above spectral density $v_s(\omega)$ can not be analytically continued
from the real axis neither to the upper nor to the lower halfplane,
thus we can not directly use the technique of generalized
spectral decomposition elaborated in [28] for the Friedrichs model
to the discrete Stark Hamiltonian under consideration.

In the conclusion let us notice that the natural continuation of the
present study would be the construction of a generalized spectral
decomposition for the extended discrete Stark Hamiltonian in a
rigged Hilbert space
$\Phi_{\pm} \subset \widehat{\cal H} \subset \Phi^{\dagger}_{\pm}$
and the proof of the weak completeness there. On this basis one can split
the unitary evolution group into two semigroups and study the problem
of intrinsic irreversibility. However it seams that the solution of these
problems should compose the subject of a separate paper.

\vskip1cm

{\large \bf Acknowledgements}

This work was supported by the Comission of the European
Communities in the frame of the EC-Russia collaboration (contract
ESPRIT P9282 ACTCS). Yu.K. gratefully
acknowledges the hospitality of the International
Solvay Institutes for Physics and Chemistry and personally
I.Prigogine for encouragement. We are thankful to I.Antoniou for
interesting discussions.
This work was partially
supported by the Russian Foundation for Basic Research under grant
\# 95-01-00568 and by the ISF under grant \# NX1300.


\newpage


\baselineskip=18pt

\begin{thebibliography}{28}

\bibitem{Avr1} Avron J and Herbst I W 1977
 {\it Comm.Math.Phys.} {\bf 52} 239

\bibitem{Avr2} Avron J E, Zak J, Grossmann A and Guuther L 1977
{\it J.Math.Phys.} {\bf 18} 918

\bibitem{Her} Herbst I W and Howland J S 1981
{\it Comm.Math.Phys.} {\bf 80} 23

\bibitem{Avr3} Avron J.E. 1982
{\it Ann.Phys.} {\bf 143} 33

\bibitem{Bus} Buslaev V S and Dmitrieva L A 1990
{\it Leningrad Math.J.} {\bf 1} 287

\bibitem{Com} Combes J M and Hislop P D 1991
{\it Comm.Math.Phys.} {\bf 140} 291

\bibitem{Ben1} Bentosela F and Grecchi V 1991
{\it Comm.Math.Phys.} {\bf 1991} 169

\bibitem{Gre1} Grecchi V, Maioli M and Saccheti A 1993
{\it J.Phys. A.} {\bf 26} L379

\bibitem{Gre2} Grecchi V, Maioli M and Saccheti A 1994
{\it Comm.Math.Phys.} {\bf 159} 605

\bibitem{Avr4} Avron J E, Exner P and Last Y 1994
{\it Phys.Rev.Lett.} {\bf 72} 896

\bibitem{Mai} Maioli M and Saccheti A 1995
{\it J.Phys. } {\bf A 28} 1101

\bibitem{Exn} Exner P.
{\it J.Math.Phys.} {\bf 36}(9) 4561

\bibitem{Ben2} Bentosela F, Carmona K, Duclos P, Simon B, Soillard B and Weder
R
1983
{\it Comm. Math. Phys.} {\bf 88} 387

\bibitem{Din} Dinaburg E I 1989
{\it Teor. i Mat. Fiz.} {\bf 78} 70

\bibitem{Lee} Lee T D 1956
{\it Phys. Rev.} {\bf 95} 1329

\bibitem{Fri} Friedrichs K O 1950
{\it Comm.Pure Appl.Math.} {\bf 1} 361

\bibitem{Pav} Pavlov B S 1987
{\it Russ. Math. Surv.} {\bf 42} 127

\bibitem{Kup} Kuperin Yu A and Merkuriev S P 1992
{\it Am. Math. Soc. Transl.} {\bf 150} 141

\bibitem{Ant} Antoniou I, Levitan J and Horwitz L P
1993
{\it J. Phys.} {\bf A 26} 6033

\bibitem{DSh} Dunford N and Schwartz J T 1988
 {\it Linear Operators. Part II:
Spectral Theory. Self Adjoint Operators in Hilbert Space}
(New York: Wiley)

\bibitem{BEl} Br\"ndas E and Elander N (Eds.) 1989
 {\it Resonances, Lecture Notes
in Physics} {\bf 325}
(Berlin Heidelberg: Springer-Verlag)

\bibitem{Ant1} Antoniou I and Tasaki S 1993
{\it Int.J.of Quantum Chemistry} {\bf 46} 425

\bibitem{Ant2} Antoniou I and Prigogine I 1993
{\it Physica A } {\bf 192} 443.

\bibitem{Bohm} Bohm A, Gadella M 1989
 {\it Dirac Kets, Gamow Vectors and Gelfand
Triplets, Lecture Notes in Physics} {\bf 348}  (Berlin: Springer-Verlag)

\bibitem{Sud} Sudarshan E C G, Chiu C and Gorini V 1978
{\it Phys.Rev.} { \bf D 18} 2914

\bibitem{Haan} de Haan M and Henin F 1973
{\it Physica} {\bf 67} 197

\bibitem{Pav1} Pavlov B S 1993
{\it St.Petersburg Math.J.} {\bf 4} 1245

\bibitem{AKDM} Antoniou I, Dmitrieva L, Kuperin Yu and
 Melnikov Yu 1995 {\it Resonances
and the Extension of Dynamics to Rigged Hilbert Space} (Preprint IPRT 98-95);
{\it Int.J.Comput.Math.Appl.}(to appear)

\bibitem{Korn} Korn G A and Korn M S 1968 {\it Mathematical Handbook}
(New York: McGraw-Hill Book Company)

\bibitem{BZP} Baz A I, Zel'dovich Ya B and Perelomov A M 1971
{\it Scattering, Reactions and Decay in Nonrelativistic Quantum
Mechanics} (Moscow: Nauka, 2nd ed. [English translation of 1st ed.:
1966 Jerusalem: Israel Program for Sientific Translations])


\end{thebibliography}
\end{document}